\documentstyle[12pt,epsf]{article}

\begin{document}


\pagestyle{empty}

\renewcommand{\thefootnote}{\fnsymbol{footnote}}
                                                  

\begin{flushright}
{\small
SLAC--PUB--8701\\
November 2000\\}
\end{flushright}
                
\begin{center}
{\bf \Large $B_s$ Mixing at SLD\footnote{Work supported in part by the
Department of Energy contract  DE--AC03--76SF00515.}}

\bigskip
\bigskip
Cheng-Ju S. Lin \\
Department of Physics, University of Massachusetts at Amherst \\
Amherst, MA 01003

\bigskip
\bigskip
Representing the SLD Collaboration$^{**}$ \\
\smallskip
Stanford Linear Accelerator Center, \\
Stanford University, Stanford, CA 94309\\
\medskip

\vspace{1.5cm}

{\bf\large
Abstract }
\end{center}
\noindent
We set a preliminary 95\% C.L. exclusion on the oscillation
frequency of $B_s^0 - \overline{B_s^0}$ mixing using a sample of 
400,000 hadronic $Z^0$ decays collected by the SLD experiment at
the SLC during the 1996-98 run. Three analyses are presented in
this paper.  The first analysis partially reconstructs the $B_s^0$ by
combining a fully reconstructed $D_s$ with the remaining charged B decay 
tracks.  The second analysis selects a sample of events with a partially
reconstructed charm vertex and a lepton track.  The third analysis
reconstructs b-hadrons topologically and exploits the
$b \rightarrow c$ cascade charge structure to determine the flavor of 
the b-hadron at decay.  All three analyses take advantage of the large 
forward-backward asymmetry of the polarized $Z^0 \rightarrow b \overline{b}$
decays and information in the hemisphere opposite to the reconstructed B vertex
to determine the b-hadron flavor at production.  The results of the three
analyses are combined to exclude the following values of the 
$B_s^0 - \overline{B_s^0}$ oscillation frequency: $\Delta m_s < 7.6 \ ps^{-1}$
and $11.8 < \Delta m_s < 14.8 \ ps^{-1}$ at the 95\% confidence level.

\vspace{2cm}

\begin{center}

{\sl Contributed to the Proceedings of DPF2000: The Meeting of the
Division of Particles and Fields of the American Physical Society,
Columbus, Ohio, 9-12 Aug 2000.}

\end{center}
\vfill

\normalsize

\pagebreak
\pagestyle{plain}

\pagebreak

\section{Introduction}
The Standard Model allows $B^0 \leftrightarrow \overline{B^0}$ oscillations
to occur via second order weak interactions.  The frequency of
oscillation is determined by the mass differences, $\Delta m$, 
between the mass eigenstates in the $B^0$ system.
The mass difference in the $B^0_s$ system ($\Delta m_s$) and 
in the $B^0_d$ system ($\Delta m_d$)
are proportional to the Cabibbo-Kobayashi-Maskawa (CKM) matrix elements  
$\left| V_{ts} \right|^2$ and $\left| V_{td} \right|^2$, respectively.
A measurement of $\Delta m_d$ can in principle be used to extract the
CKM matrix element $\left| V_{td} \right|$.
However, the extraction of $\left| V_{td} \right|$ from $\Delta m_d$ 
is complicated by a large theoretical uncertainty on the hadronic
matrix elements.  The complication can be circumvented by taking the ratio
of $\Delta m_s$ and $\Delta m_d$.  In the ratio, the theoretical uncertainty
is reduced to the 5\% level [1].
Therefore, a direct measurement of $\Delta m_s$, combined with the
current measurement of $\Delta m_d$, can be translated to 
a precise value of $\left| V_{td} \right|$.

\section{Experimental Technique}
The results presented in this paper are based on 400,000 hadronic
$Z^0$ decays collected during the 1996-98 run with an average electron
beam polarization of 73\%.  A detailed description of the experimental
apparatus can be found elsewhere [2].  
The three main ingredients for measuring the time dependent 
$B_s^0 - \overline{B_s^0}$ 
oscillations are: (1) determination of the flavor at production 
(initial state tag), (2) determination of the flavor at the decay 
vertex (final state tag), and
(3) reconstruction of the proper decay time of the $B_s^0$.
Three methods were explored at SLD for studying the $B_s^0$ oscillations 
and the methods are referred to as ``$D_s$+Tracks'', ``Lepton+D'', and 
``Charge Dipole'' [3,4].
All three analyses share the common initial state tag as well as the B energy
reconstruction algorithms but differ in the event selection,
decay length reconstruction, and the final state tag.  

Several techniques are used to determine the initial 
state of the $B_s^0$.  The most powerful method, unique to the SLD, 
is the polarization tag.  In a 
polarized $Z^0 \rightarrow b\overline{b}$ decay, the 
outgoing quark is produced preferentially along the direction 
opposite to the spin of the $Z^0$ boson.  Therefore by 
knowing the helicity of the electron beam and the 
direction of the jet, the flavor of the primary quark 
in the jet can be determined.  To further enhance the initial
state tag, information in the opposite hemisphere (e.g.
momentum weighted jet charge, vertex charge, lepton and
kaon tracks) is used.  Combining all available tags, the
average initial state correct tag probability is about
78\%.

An ideal mixing analysis requires high efficiency, high $B_s^0$ purity,
clean initial and final state tags,  and
excellent proper time resolution.  In practice, experimental constraints
necessitate trade-offs between the four key elements.  
The ``$D_s$+tracks'' is the most exclusive
analysis at SLD.  The analysis partially reconstructs the
$B_s^0$ by combining a fully reconstructed $D_s$ 
(via $\phi \pi$ and $K^{*0}K$ modes) with other secondary B decay tracks.
By taking the exclusive approach, the analysis is able to achieve
a high average $B_s^0$ purity of 38\% and an excellent decay
length resolution of 48$\mu m$ (60\% core resolution) for
the $B_s^0$ events.  However, the analysis suffers from low efficiency and 
only 361 candidates are selected in the final sample.  
The ``Lepton+D'' takes a slightly
more inclusive approach by selecting events with a partially reconstructed
D vertex and a lepton in the same hemisphere.  The identified lepton
not only enhances the b-hadron fraction but also provides a clean
final state tag (final state mistag $<$ 10\%).  
The estimated B energy resolution is comparable to the other two
analyses and is about 7\% (60\% core fraction) and 20\% (tail) for the
$B_s^0 (b \rightarrow {\it l})$ events. 
The ``Lepton+D'' has
2087 candidates in the final sample with an average $B_s^0$ purity 
of about 16\%.  The most inclusive method and also the analysis with 
the highest sensitivity at SLD is the ``Charge Dipole''.  The ``Charge
Dipole'' selects events that contain both a secondary and a
tertiary vertex.  To enhance the $B_s^0$ fraction, the total
track charge (from secondary and tertiary) is required to be zero.
The final state tag is based on the dipole value which is defined as the
sign of the charge difference weighted by the distance between
the secondary and the tertiary vertices.  The final state correct tag
probability using the dipole method is highly dependent on the decay
toplogy and ranges from 53\% for $B_s^0 \rightarrow D \overline{D} X$ 
to 91\% for $B_s^0 \rightarrow D_s X$ decays.
A total of 8556 decays is
selected in the ``Charge Dipole'' analysis with a $B_s^0$ purity 
of 15\%.

\section{Results}
The study of the $B_s^0$ oscillations is performed using the 
amplitude fit method [5].
In this method, the probability for mixing, which is proportional to
$1-cos(\Delta m_s \tau)$, is modified by introducing the amplitude
{\bf A} in front of the cosine (same modification for the unmixed
expression).
The amplitude plot is generated by scanning the $\Delta m_s$ value over
a specified range, and for each $\Delta m_s$ value, fitting for the 
parameter {\bf A}.
We expect the fitted value of {\bf A} to
be consistent with zero when the chosen $\Delta m_s$ is away
from the true value and the amplitude {\bf A} to reach the
value 1 near the true $\Delta m_s$.  If no signal is seen,
a 95$\%$ C.L. lower limit can be set for frequencies at which
${\bf A} + 1.645\sigma_A < 1$.  The 95$\%$ C.L.
sensitivity is defined as the value of $\Delta m_s$ at which 
$1.645\sigma_A=1$.

The amplitude plot for the three SLD analyses combined 
is shown in Fig. 1.
The combined plot takes into account the correlated systematic uncertainties.
Furthermore, the samples were selected such as to remove any statistical
overlap between analyses.
No evidence of a signal is observed up to $\Delta m_s$ of 25 $ps^{-1}$.
The preliminary SLD results exclude the following values 
of $\Delta m_s$ at the 
95\% C.L.: $\Delta m_s $ $<$ 7.6 $ps^{-1}$ and 
11.8 $<$ $\Delta m_s$ $<$ 14.8 $ps^{-1}$.  The SLD combined sensitivity at 
the 95\% C.L. is 13.0 $ps^{-1}$.

\begin{figure}[htb]
\begin{center}{
\epsfxsize15cm
\epsfysize10cm
\epsfbox{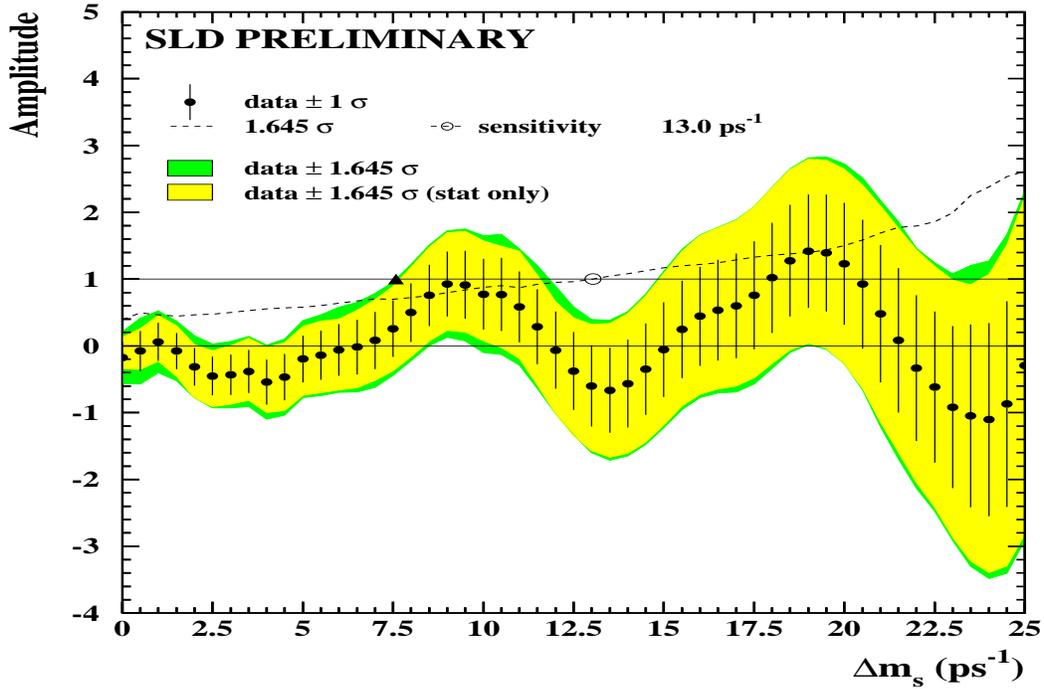}
\caption{SLD preliminary amplitude plot.}}
\end{center}
\end{figure}

\section*{Acknowledgments}
We thank the personnel of the SLAC accelerator department and
the technical staffs of our collaborating institutions for their outstanding
efforts.  This work was supported by the Department of Energy, the National
Science Foundataion, the Instituto Nazionale di Fisica of Italy, the
Japan-US Cooperative Research Project on High Energy Physics, and the
Science and Engineering Research Council of the United Kingdom.

\newpage
%
%
%
\section*{$^{**}$ List of Authors}

\begin{center}
\def\iAOMORI{$^{(1)}$}
\def\iBRI{$^{(2)}$}
\def\iBRUN{$^{(3)}$}
\def\iBU{$^{(4)}$}
\def\iCOLO{$^{(5)}$}
\def\iCSU{$^{(6)}$}
\def\iFERR{$^{(7)}$}
\def\iFRAS{$^{(8)}$}
\def\iJHU{$^{(9)}$}
\def\iLBL{$^{(10)}$}
\def\iMASS{$^{(11)}$}
\def\iMISSI{$^{(12)}$}
\def\iMIT{$^{(13)}$}
\def\iMOSCOW{$^{(14)}$}
\def\iNAGO{$^{(15)}$}
\def\iOREG{$^{(16)}$}
\def\iOXF{$^{(17)}$}
\def\iPERU{$^{(18)}$}
\def\iRAL{$^{(19)}$}
\def\iRUTG{$^{(20)}$}
\def\iSLAC{$^{(21)}$}
\def\iSOONG{$^{(22)}$}
\def\iTENN{$^{(23)}$}
\def\iTOHO{$^{(24)}$}
\def\iUCSB{$^{(25)}$}
\def\iUCSC{$^{(26)}$}
\def\iVAND{$^{(27)}$}
\def\iWASH{$^{(28)}$}
\def\iWISC{$^{(29)}$}
\def\iYALE{$^{(30)}$}

  \baselineskip=.75\baselineskip  
\mbox{Kenji Abe\unskip,\iNAGO}
\mbox{Koya Abe\unskip,\iTOHO}
\mbox{T. Abe\unskip,\iSLAC}
\mbox{I. Adam\unskip,\iSLAC}
\mbox{H. Akimoto\unskip,\iSLAC}
\mbox{D. Aston\unskip,\iSLAC}
\mbox{K.G. Baird\unskip,\iMASS}
\mbox{C. Baltay\unskip,\iYALE}
\mbox{H.R. Band\unskip,\iWISC}
\mbox{T.L. Barklow\unskip,\iSLAC}
\mbox{J.M. Bauer\unskip,\iMISSI}
\mbox{G. Bellodi\unskip,\iOXF}
\mbox{R. Berger\unskip,\iSLAC}
\mbox{G. Blaylock\unskip,\iMASS}
\mbox{J.R. Bogart\unskip,\iSLAC}
\mbox{G.R. Bower\unskip,\iSLAC}
\mbox{J.E. Brau\unskip,\iOREG}
\mbox{M. Breidenbach\unskip,\iSLAC}
\mbox{W.M. Bugg\unskip,\iTENN}
\mbox{D. Burke\unskip,\iSLAC}
\mbox{T.H. Burnett\unskip,\iWASH}
\mbox{P.N. Burrows\unskip,\iOXF}
\mbox{A. Calcaterra\unskip,\iFRAS}
\mbox{R. Cassell\unskip,\iSLAC}
\mbox{A. Chou\unskip,\iSLAC}
\mbox{H.O. Cohn\unskip,\iTENN}
\mbox{J.A. Coller\unskip,\iBU}
\mbox{M.R. Convery\unskip,\iSLAC}
\mbox{V. Cook\unskip,\iWASH}
\mbox{R.F. Cowan\unskip,\iMIT}
\mbox{G. Crawford\unskip,\iSLAC}
\mbox{C.J.S. Damerell\unskip,\iRAL}
\mbox{M. Daoudi\unskip,\iSLAC}
\mbox{S. Dasu\unskip,\iWISC}
\mbox{N. de Groot\unskip,\iBRI}
\mbox{R. de Sangro\unskip,\iFRAS}
\mbox{D.N. Dong\unskip,\iSLAC}
\mbox{M. Doser\unskip,\iSLAC}
\mbox{R. Dubois\unskip,\iSLAC}
\mbox{I. Erofeeva\unskip,\iMOSCOW}
\mbox{V. Eschenburg\unskip,\iMISSI}
\mbox{S. Fahey\unskip,\iCOLO}
\mbox{D. Falciai\unskip,\iFRAS}
\mbox{J.P. Fernandez\unskip,\iUCSC}
\mbox{K. Flood\unskip,\iMASS}
\mbox{R. Frey\unskip,\iOREG}
\mbox{E.L. Hart\unskip,\iTENN}
\mbox{K. Hasuko\unskip,\iTOHO}
\mbox{S.S. Hertzbach\unskip,\iMASS}
\mbox{M.E. Huffer\unskip,\iSLAC}
\mbox{X. Huynh\unskip,\iSLAC}
\mbox{M. Iwasaki\unskip,\iOREG}
\mbox{D.J. Jackson\unskip,\iRAL}
\mbox{P. Jacques\unskip,\iRUTG}
\mbox{J.A. Jaros\unskip,\iSLAC}
\mbox{Z.Y. Jiang\unskip,\iSLAC}
\mbox{A.S. Johnson\unskip,\iSLAC}
\mbox{J.R. Johnson\unskip,\iWISC}
\mbox{R. Kajikawa\unskip,\iNAGO}
\mbox{M. Kalelkar\unskip,\iRUTG}
\mbox{H.J. Kang\unskip,\iRUTG}
\mbox{R.R. Kofler\unskip,\iMASS}
\mbox{R.S. Kroeger\unskip,\iMISSI}
\mbox{M. Langston\unskip,\iOREG}
\mbox{D.W.G. Leith\unskip,\iSLAC}
\mbox{V. Lia\unskip,\iMIT}
\mbox{C. Lin\unskip,\iMASS}
\mbox{G. Mancinelli\unskip,\iRUTG}
\mbox{S. Manly\unskip,\iYALE}
\mbox{G. Mantovani\unskip,\iPERU}
\mbox{T.W. Markiewicz\unskip,\iSLAC}
\mbox{T. Maruyama\unskip,\iSLAC}
\mbox{A.K. McKemey\unskip,\iBRUN}
\mbox{R. Messner\unskip,\iSLAC}
\mbox{K.C. Moffeit\unskip,\iSLAC}
\mbox{T.B. Moore\unskip,\iYALE}
\mbox{M. Morii\unskip,\iSLAC}
\mbox{D. Muller\unskip,\iSLAC}
\mbox{V. Murzin\unskip,\iMOSCOW}
\mbox{S. Narita\unskip,\iTOHO}
\mbox{U. Nauenberg\unskip,\iCOLO}
\mbox{H. Neal\unskip,\iYALE}
\mbox{G. Nesom\unskip,\iOXF}
\mbox{N. Oishi\unskip,\iNAGO}
\mbox{D. Onoprienko\unskip,\iTENN}
\mbox{L.S. Osborne\unskip,\iMIT}
\mbox{R.S. Panvini\unskip,\iVAND}
\mbox{C.H. Park\unskip,\iSOONG}
\mbox{I. Peruzzi\unskip,\iFRAS}
\mbox{M. Piccolo\unskip,\iFRAS}
\mbox{L. Piemontese\unskip,\iFERR}
\mbox{R.J. Plano\unskip,\iRUTG}
\mbox{R. Prepost\unskip,\iWISC}
\mbox{C.Y. Prescott\unskip,\iSLAC}
\mbox{B.N. Ratcliff\unskip,\iSLAC}
\mbox{J. Reidy\unskip,\iMISSI}
\mbox{P.L. Reinertsen\unskip,\iUCSC}
\mbox{L.S. Rochester\unskip,\iSLAC}
\mbox{P.C. Rowson\unskip,\iSLAC}
\mbox{J.J. Russell\unskip,\iSLAC}
\mbox{O.H. Saxton\unskip,\iSLAC}
\mbox{T. Schalk\unskip,\iUCSC}
\mbox{B.A. Schumm\unskip,\iUCSC}
\mbox{J. Schwiening\unskip,\iSLAC}
\mbox{V.V. Serbo\unskip,\iSLAC}
\mbox{G. Shapiro\unskip,\iLBL}
\mbox{N.B. Sinev\unskip,\iOREG}
\mbox{J.A. Snyder\unskip,\iYALE}
\mbox{H. Staengle\unskip,\iCSU}
\mbox{A. Stahl\unskip,\iSLAC}
\mbox{P. Stamer\unskip,\iRUTG}
\mbox{H. Steiner\unskip,\iLBL}
\mbox{D. Su\unskip,\iSLAC}
\mbox{F. Suekane\unskip,\iTOHO}
\mbox{A. Sugiyama\unskip,\iNAGO}
\mbox{S. Suzuki\unskip,\iNAGO}
\mbox{M. Swartz\unskip,\iJHU}
\mbox{F.E. Taylor\unskip,\iMIT}
\mbox{J. Thom\unskip,\iSLAC}
\mbox{E. Torrence\unskip,\iMIT}
\mbox{T. Usher\unskip,\iSLAC}
\mbox{J. Va'vra\unskip,\iSLAC}
\mbox{R. Verdier\unskip,\iMIT}
\mbox{D.L. Wagner\unskip,\iCOLO}
\mbox{A.P. Waite\unskip,\iSLAC}
\mbox{S. Walston\unskip,\iOREG}
\mbox{A.W. Weidemann\unskip,\iTENN}
\mbox{E.R. Weiss\unskip,\iWASH}
\mbox{J.S. Whitaker\unskip,\iBU}
\mbox{S.H. Williams\unskip,\iSLAC}
\mbox{S. Willocq\unskip,\iMASS}
\mbox{R.J. Wilson\unskip,\iCSU}
\mbox{W.J. Wisniewski\unskip,\iSLAC}
\mbox{J.L. Wittlin\unskip,\iMASS}
\mbox{M. Woods\unskip,\iSLAC}
\mbox{T.R. Wright\unskip,\iWISC}
\mbox{R.K. Yamamoto\unskip,\iMIT}
\mbox{J. Yashima\unskip,\iTOHO}
\mbox{S.J. Yellin\unskip,\iUCSB}
\mbox{C.C. Young\unskip,\iSLAC}
\mbox{H. Yuta\unskip.\iAOMORI}

\it
  \vskip \baselineskip                   
  \centerline{(The SLD Collaboration)}   
  \vskip \baselineskip        
  \baselineskip=.75\baselineskip   
\iAOMORI
  Aomori University, Aomori , 030 Japan, \break
\iBRI
  University of Bristol, Bristol, United Kingdom, \break
\iBRUN
  Brunel University, Uxbridge, Middlesex, UB8 3PH United Kingdom, \break
\iBU
  Boston University, Boston, Massachusetts 02215, \break
\iCOLO
  University of Colorado, Boulder, Colorado 80309, \break
\iCSU
  Colorado State University, Ft. Collins, Colorado 80523, \break
\iFERR
  INFN Sezione di Ferrara and Universita di Ferrara, I-44100 Ferrara, Italy, \break
\iFRAS
  INFN Lab. Nazionali di Frascati, I-00044 Frascati, Italy, \break
\iJHU
  Johns Hopkins University,  Baltimore, Maryland 21218-2686, \break
\iLBL
  Lawrence Berkeley Laboratory, University of California, Berkeley, California 94720, \break
\iMASS
  University of Massachusetts, Amherst, Massachusetts 01003, \break
\iMISSI
  University of Mississippi, University, Mississippi 38677, \break
\iMIT
  Massachusetts Institute of Technology, Cambridge, Massachusetts 02139, \break
\iMOSCOW
  Institute of Nuclear Physics, Moscow State University, 119899, Moscow Russia, \break
\iNAGO
  Nagoya University, Chikusa-ku, Nagoya, 464 Japan, \break
\iOREG
  University of Oregon, Eugene, Oregon 97403, \break
\iOXF
  Oxford University, Oxford, OX1 3RH, United Kingdom, \break
\iPERU
  INFN Sezione di Perugia and Universita di Perugia, I-06100 Perugia, Italy, \break
\iRAL
  Rutherford Appleton Laboratory, Chilton, Didcot, Oxon OX11 0QX United Kingdom, \break
\iRUTG
  Rutgers University, Piscataway, New Jersey 08855, \break
\iSLAC
  Stanford Linear Accelerator Center, Stanford University, Stanford, California 94309, \break
\iSOONG
  Soongsil University, Seoul, Korea 156-743, \break
\iTENN
  University of Tennessee, Knoxville, Tennessee 37996, \break
\iTOHO
  Tohoku University, Sendai 980, Japan, \break
\iUCSB
  University of California at Santa Barbara, Santa Barbara, California 93106, \break
\iUCSC
  University of California at Santa Cruz, Santa Cruz, California 95064, \break
\iVAND
  Vanderbilt University, Nashville,Tennessee 37235, \break
\iWASH
  University of Washington, Seattle, Washington 98105, \break
\iWISC
  University of Wisconsin, Madison,Wisconsin 53706, \break
\iYALE
  Yale University, New Haven, Connecticut 06511. \break

\rm
%

\end{center}

\enddocument